# Photodynamics and Temperature Dependence of Single Spin Defects in Hexagonal Boron Nitride

Benjamin Whitefield[†,1,2], Ivan Zhigulin[†,1,2], Nicholas P. Sloane[1,2], Jean-Philippe Tetienne[3], Igor Aharonovich[*,1,2] and Mehran Kianinia[*,1,2]

[1] School of Mathematical and Physical Sciences, University of Technology Sydney, Ultimo, New South Wales 2007, Australia
[2] ARC Centre of Excellence for Transformative Meta-Optical Systems, University of Technology Sydney, Ultimo, New South Wales 2007, Australia
[3] Department of Physics, School of Science, RMIT University, Melbourne, VIC 3001, Australia

† These authors contributed equally to this work.
* To whom correspondence should be addressed: I.A. Igor.Aharonovich@uts.edu.au, M.K. Mehran.Kianinia@uts.edu.au

**ABSTRACT**

*Quantum emitters in hexagonal boron nitride (hBN) that exhibit optically detected magnetic resonance (ODMR) signatures have recently garnered significant attention as an emerging solid-state platform for quantum technologies. However, the underlying spin dynamics, and the mechanisms determining the spin-dependent fluorescence in these defects are still poorly understood. In this work we perform detailed photodynamical studies of the spin complexes in hBN. In particular, we show that spin transitions are located within the metastable manifold which can be explained by the rate model, populating in a cascading manner. In addition, we perform temperature dependent measurements on these defects and show that the spin-lattice relaxation and coherence times increase as the temperature reduces. Furthermore, we find that the ODMR frequencies of the S=1 transition show only a marginal frequency shift as a function of temperature, which makes them a robust sensor at cryogenic temperatures. These insights are crucial for further understanding of the spin dynamics of quantum emitters in hBN and their practical implementation in quantum sensing.*

**INTRODUCTION**

Solid state defects have become prominent building blocks for emerging quantum technologies, including sensing, communication and information processing[1–3]. Many of these advances hinge on the development of quantum emitters (QEs) featuring optically addressable spins[4–7]. Hexagonal boron nitride (hBN) has established itself as a notable platform for exploring spin-active defects, offering a two-dimensional structure with distinct optical and spin properties[7–12]. hBN complements traditional three-dimensional solid state hosts by enabling access to new spin phenomena and facilitating different avenues for scalable quantum photonic devices[13–15].

The negatively charged boron vacancy ($V_B^-$) has long been the focus of spin defect studies in hBN, serving as a room temperature quantum sensor compatible with integrated quantum device architectures[16–21]. Despite its attractive properties, the low quantum efficiency of $V_B^-$ and the inability to isolate single defects have impeded its performance. These challenges have driven interest in emerging single QEs in hBN, which provide individually addressable spins[7,10,22–25]. In recent years, several studies have reported single spin-active defects in hBN. While some have observed only S = 1 manifolds[9,25], most have described an S = ½ -like spin signature that has been attributed to a weakly coupled electron spin-pair and have also been observed in solid state materials[11,26–28].

Recently, a previously unexplored class of spin defects has emerged, commonly referred to as the spin complex. This defect acts as a QE, while featuring spin transitions linked to both the weakly coupled spin-pair, as well as those arising from an S = 1 triplet state[10,26]. The coexistence of distinct spin manifolds within a single emitter offers a compelling route towards more flexible spin control schemes, yet its fundamental nature remains poorly understood. The internal level structures, relaxation mechanisms and temperature dependencies are largely unresolved. In this work, we explore these physical principles by pursuing two complementary lines of investigation. First, by probing the internal transition rates and constructing a spin dependent rate model to accurately describe experimental data. Then, through the exploration of the temperature dependence of both optical and spin transitions as well as spin coherence times.

**RESULTS AND DISCUSSION**

Figure 1a illustrates the spin complex single photon emitters in hBN, comprising two independent defects between which charge transfer can occur. Spin complex defects can be generated by annealing carbon rich hBN at 1000°C in an oxygen environment[12]. The charge transfer between these nearby defects give rise to two spin states: When both electrons occupy one defect, they are strongly coupled, creating an S = 1 manifold, as shown in the top left box (orange) of Figure 1a. After charge transfer, the spin complex enters a weakly coupled regime, as described in the right hand box(red). This spin configuration is characterized by a mixture of singlet and triplet states, S = {0,1}. A photoluminescence (PL) spectrum of an example spin complex emitter in hBN is presented in Figure 1b with a zero phonon line (ZPL) centered on ~720 nm. A second order autocorrelation of the same emitter with $g^{(2)}(0)$ of 0.24 without any background correction is shown in the inset of Figure 1b. The low value of $g^{(2)}(0)$ suggests that only one of the defects is optically active within the measurement window (700-1000 nm).

Figure 1c exhibits a characteristic ODMR spectrum from the spin complex emitter, comprising a central peak arising from S = {0,1} and two accompanying peaks within the S = 1 transitions. Note that this measurement is recorded in the presence of ~ 68 mT magnetic field applied out–of–plane to the hBN flake. A 100 um copper wire placed near the flakes carries RF which excites spin transitions, giving rise to change in PL contrast in the ODMR measurement. The transition attributed to the S = {0,1} sublevels is commonly referred as -½ ↔ +½, and will be described as such hereafter. In Figure 1c the -½ ↔ +½ transition highlighted in red occurs at 1.9 GHz while the 0 ↔ -1 and 0 ↔ +1 transitions highlighted in orange appear at 1 and 2.7 GHz, respectively. The time-resolved PL decay of a typical spin complex emitter is shown in Figure 1d and is fit with a single exponential. This reveals a lifetime of 2.5 ns, characteristic of most hBN quantum emitters[29].

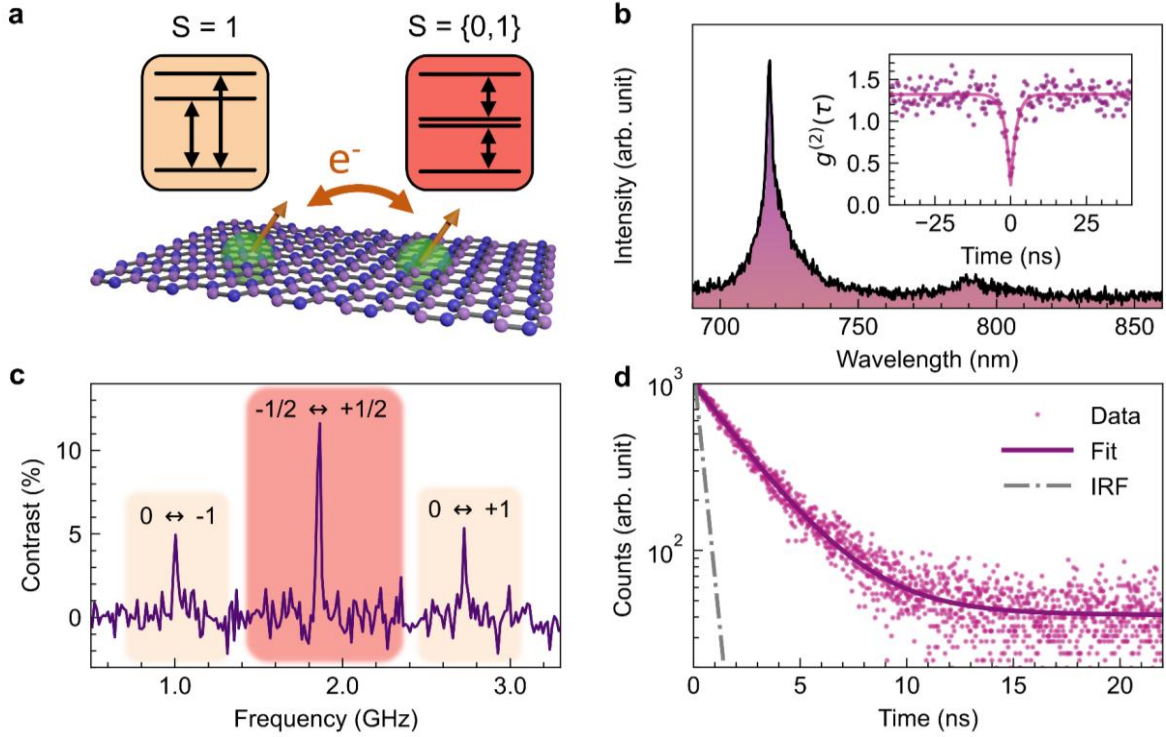

*Figure 1. **The Spin Complex.** (a) Schematic of the spin complex, consisting of two defects in the hBN lattice with charge transfer occurring between them. The spin manifolds describe the S = 1 (orange) and S = {0,1} transitions (red). (b) An example photoluminescence spectrum of an ODMR-active single photon emitter in hBN. Inset is the second order autocorrelation function demonstrating single photon emission with a $g^{(2)}(0)$ of 0.24. (c) ODMR spectrum of the emitter shown in (b), the 0 ↔ -1 (1 GHz) and 0 ↔ +1 (2.7 GHz) transitions are highlighted in orange and the -½ ↔ +½ transition (1.9 GHz) is highlighted in red. (d) Time-resolved photoluminescence decay which is fit with a single exponential to reveal a fluorescent lifetime of 2.5 ns. The instrument response function (IRF), shown in grey, is much shorter than the fluorescent lifetime.*

A detailed understanding of the spin complex requires analysis of the internal electron transition rates that dictate population dynamics throughout optical excitation, dark intervals and spin manipulation. We therefore begin by investigating the spin-dependent relaxation pathways through pulsed ODMR measurements. Figure 2a illustrates the pulse sequence for coherent driving of various spin transitions which results in rabi oscillations. The system is first initialized via optical pumping, then a resonant RF pulse of varying length ($\tau$) drives the chosen spin transition. A final laser pulse is applied for PL readout of the resulting spin state. The PL contrast is calculated based on mean values of photon counts in the signal and reference, denoted by the purple and pink dashed rectangles. Rabi oscillations were driven on both the 0 ↔ +1 and -½ ↔ +½ spin transitions (Figure 2c). Further characterization of this emitter can be seen in Supplementary Information Figure 1. Both transitions exhibited coherent driving, with corresponding $\pi$-pulse durations of 42 and 87 ns, respectively. Importantly, the Rabi oscillations show distinct damping rates while the envelope amplitude increases with $\tau$, implying a long spin dependent recovery process occurring under resonant driving. This long recovery time indicates the presence of long lived metastable states and its presence in the rabi oscillation implies that spin states occupy metastable states, as will be discussed in more detail later. To further explore these features, the oscillations were fitted with a damped cosine with an exponential component to account for the long recovery rate:

$$A * cos(2\pi\nu\tau - \phi) * e^{-k_1\tau} + B * (1 - e^{-k_2\tau}) \quad (1)$$

where $\nu$ is the Rabi frequency, $\phi$ is the phase and A is the amplitude. B is the amplitude of the depopulation rate, while $k_1$ and $k_2$ are the rate constants for the damping and recovery time (from metastable state), respectively. The damping of the Rabi oscillation envelope reflects contributions from both homogeneous ($T_2$) and inhomogeneous ($T_2^*$) dephasing mechanisms. In this spin complex, the damping time ($1/k_1$) for the 0 ↔ +1 spin transition is 147 ns whereas the -½ ↔ +½ transition exhibits substantially faster damping at 75 ns. The heightened damping in the -½ ↔ +½ transition is attributed to the distinct local environments experienced by the two electrons in the weakly coupled spin pair configuration. The metastable state depopulation time constants ($1/k_2$) are similar for the 0 ↔ +1 and -½ ↔ +½ spin transitions at 190 and 212 ns respectively.

One approach to probe the internal transitions rates of a spin active emitter is to analyze the ODMR contrast as a function of readout gate timing. By shifting the readout gate, photons emitted after different delays can be selectively measured, capturing the evolution of the spin population as metastable state electrons relax into the ground state. Investigation of this effect was carried out using the pulse sequence illustrated in Figure 2b. The sequences consist of an initialization and readout laser separated by a $\pi$-pulse for spin control. The gates (dashed rectangles) can be shifted relative to the laser pulses by a variable offset ($\tau$), enabling the identification of key features in the subsequent contrast plot. This pulse sequence was carried out with a $\pi$-pulse on resonance with the 0 ↔ +1 transition and a time trace of the signal and reference pulses are shown in Figure 2c. The $\pi$-pulse flips the spin to the shorter lived state prior to readout, producing an increase in fluorescence at the beginning of the readout laser. The system is then gradually re-initialized over ~10 μs by the readout laser until the fluorescence difference vanishes. The contrast is plotted against the gate offset (Figure 2d inset) as described in Figure 2b. Here, a clear increase in ODMR contrast is observed over the first microsecond before the eventual spin re-initialization and loss of contrast. This behaviour is reminiscent of spin systems with a ground state S = 1 manifold, such as $NV^-$ and $V_B^-$ centers, where contrast evolution is governed by spin dependent metastable state decay[30,31]. Likewise, systems with metastable state S = 1, such as molecular spins, exhibit a similar initial increase in contrast[32]. In all cases, the timescale of the observed contrast growth closely corresponds to the measured lifetimes of the metastable state[33–35]. The peak contrast for the spin complex measured in this work was found at ~800 ns, hinting at the presence of a relatively slow repopulation of the ground state. While the spin dependent metastable state decay appears on this measurement, it still remains unclear whether the driven spin transitions occur in the ground or metastable state.

To confirm whether the spin transitions belong to the metastable state within the spin complex emitter, $T_1$ relaxation measurements were performed in which an initialization laser pulse and $\pi$-pulse polarize the spin into the short lived state (Figure 2e inset). The system is then left dark for varying time ($\tau$) until the final spin state is read out optically. This measurement is key to distinguishing spin systems with ground state manifolds from those with metastable state manifolds. In a ground state system, ODMR contrast decreases steadily with dark time in the $T_1$ measurement. Once the electron has decayed from the metastable to ground state, the $\pi$-pulse will flip the spin state and contrast can be read out. Further increase in dark time between the $\pi$-pulse and readout laser only decreases contrast through spin-lattice relaxation ($T_1$) within the ground state manifold. This effect is independent of metastable state lifetime, since the $\pi$-pulse can flip the spin only when the electron resides in the ground state. In contrast, the same measurement on a metastable state spin system exhibits an initial increase in contrast ($T_{rise}$) before $T_1$ relaxation begins to dominate[35,36]. In this case, $T_{rise}$ reflects the difference in decay rates from the short and long lived spin sublevels within the metastable state. At a dark time below the short lived sublevel lifetime, the π-pulse flips the spin, however the electron remains in the metastable state and is unavailable for optical readout. Thus, the ODMR contrast initially rises as the short lived state decays into the ground state. Beyond $T_{rise}$, the contrast decreases as the combined effects of spin-lattice relaxation and electron decay from the long lived state. The effective spin-lattice relaxation time ($T_1$) arises from the combined contributions of these two relaxation rates.

The $T_1$ spin relaxation measurements are performed on both the 0 ↔ +1 and -½ ↔ +½ spin transitions and shown in Figure 2e,f. Here, $T_{rise}$ is readily observed, with contrast rising for the first 300 and 100 ns for 0 ↔ +1 and -½ ↔ +½ respectively. A bi-exponential fit is then performed, accounting for the two competing processes of $T_{rise}$ and $T_1$. For 0 ↔ +1, $T_{rise}$ and $T_1$ were found to be 87 and 5447 ns, respectively. Both $T_{rise}$ and $T_1$ were shorter for the -½ ↔ +½ transitions, equaling 27 and 4298 ns, respectively. The presence of $T_{rise}$ is a clear indication of a spin system residing in a metastable state. From the data, we can reasonably infer that the short lived state has a lifetime on the order of tens to hundreds of nanoseconds. However, the fit results alone do not capture the full picture. Namely, separating the contributions of spin-lattice relaxation and the long lived state lifetime requires a deeper understanding of the underlying photodynamics.

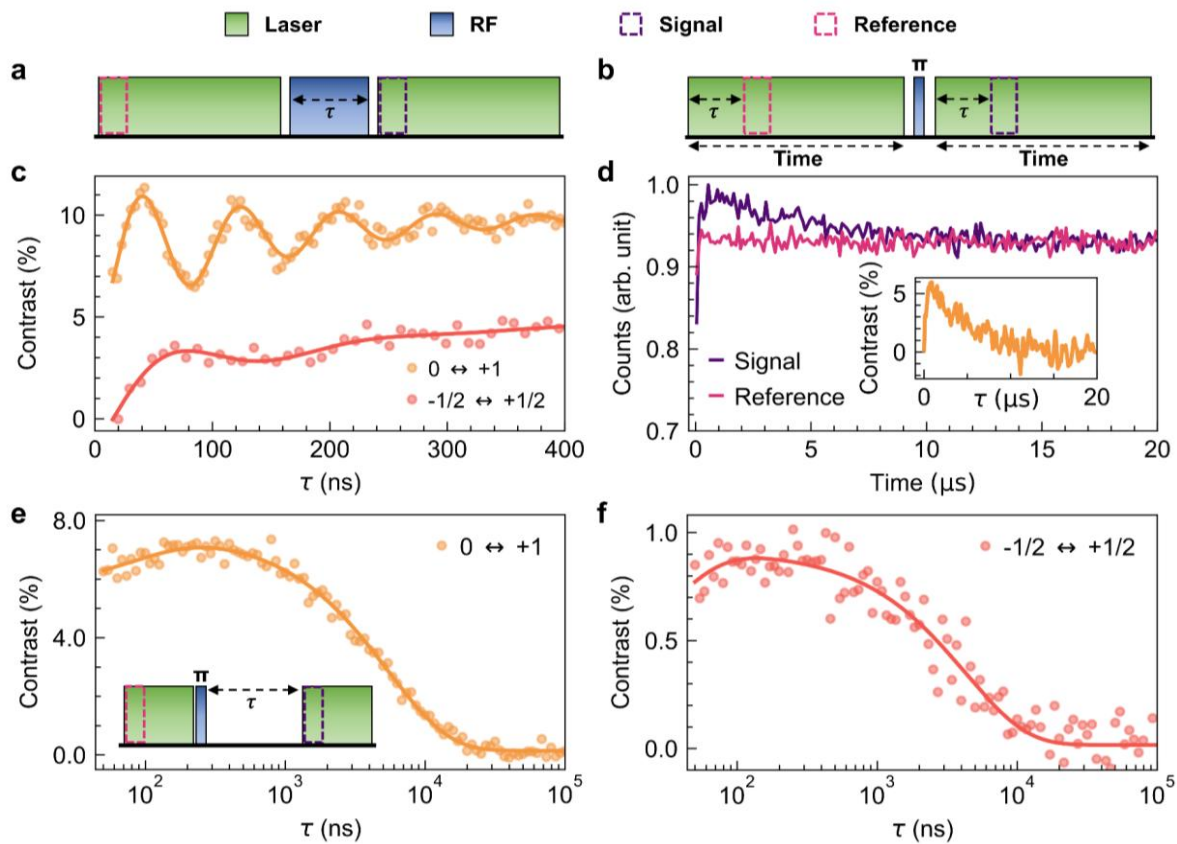

*Figure 2. Pulsed ODMR photodynamics. (a) Pulse sequence for performing Rabi oscillations. Initialization laser and readout pulses are separated by an RF pulse of increasing length (τ). The dashed rectangles depict the gated collection of the signal (purple) and reference (pink) which are used for ODMR contrast calculations. (b) Pulse sequence consisting of an initialization and readout laser pulse separated by a π-pulse. The gate offset is of varying length (τ). (c) Rabi oscillations conducted on both the 0 ↔ +1 and -½ ↔ +½ spin transitions. Experimental data is fit with a damped cosine with an additional exponential term accounting for gradual MS state depletion. (d) Time trace of both the initialization and readout laser pulses as shown in (b). Inset is the calculated contrast using a gated collection time of*

*300 ns and a gate offset of increasing length (τ). **(e)** and **(f)** show $T_1$ relaxation measurements completed on the 0 ↔ +1 and -½ ↔ +½ spin transitions respectively using the pulse sequence as shown in the inset of (e). The system is initialized into short lived state using a π-pulse, then the fluorescent response is readout after a varying delay time (τ).*

To accomplish this, a spin dependent rate model was developed, with a simplified energy level diagram presented in Figure 3a. The system is described by a singlet ground state |1⟩ and excited state |2⟩, between which optical excitation ($\kappa_p$) and radiative decay ($\kappa_r$) occur. The S = 1 and S = {0,1} spin manifolds are both associated with metastable states, consistent with earlier observations. These metastable manifolds are arranged in a cascading configuration, with population flowing from S = 1 into S = {0,1}. This configuration accounts for the suppressed ODMR contrast of the S = 1 system at zero magnetic field[12]. At low fields, the spin levels within S = {0,1} become nearly degenerate, enhancing spin mixing. As the electron passes through this manifold, spin information is lost, eliminating ODMR contrast from both manifolds.

The S = 1 triplet is placed above the S = {0,1} manifold, consistent with the shorter measured $T_{rise}$. For simplicity, the S = 1 triplet is reduced to states |3⟩ and |4⟩ (representing $m_s$ = ±1 and $m_s$ = 0), while the S = {0,1} manifold is represented by |5⟩ and |6⟩ (corresponding to $m_s$ = $T_\pm$ and $m_s$ = S,$T_0$). Intersystem crossing from the excited state occurs via rates $ISC_{23}$ and $ISC_{24}$. Metastable state decay rates $\kappa_{ij}$ describe population transfer from state i to j, while $\omega_{34}$ and $\omega_{56}$ represent the spin-lattice relaxation within each manifold. Transitions that do not conserve spin ($\kappa_{45}$ and $\kappa_{36}$) were assumed to occur at negligible rates. The driven spin transitions (0 ↔ +1 and -½ ↔ +½) are indicated by the 1 (orange) and ½ (red) connections between levels.

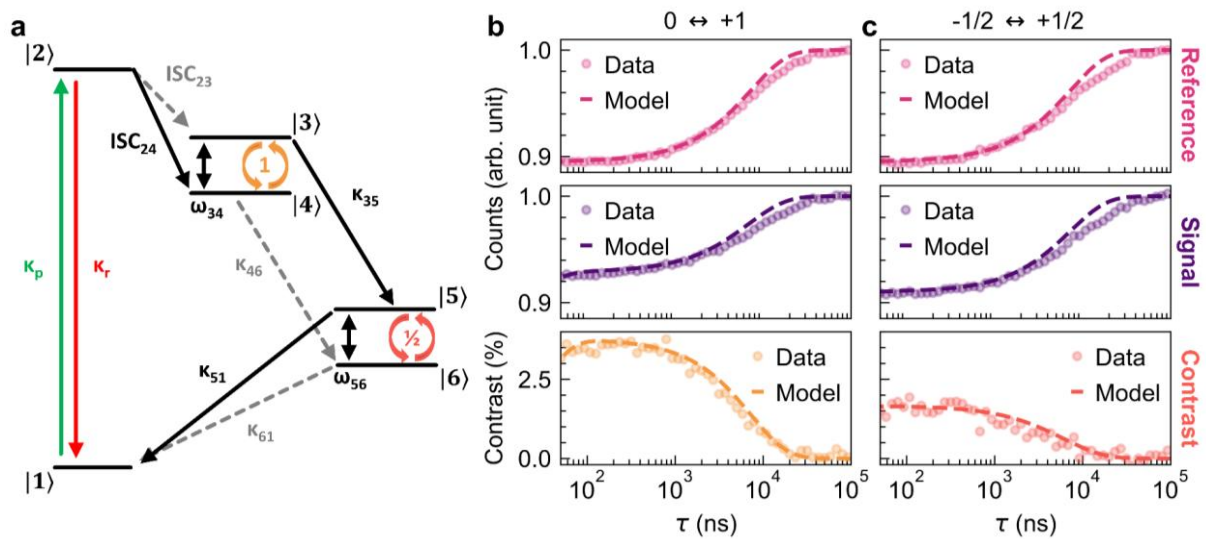

*Figure 3. Rate modelling of the spin complex. **(a)** Simplified energy level diagram of the spin complex with the various radiative and non-radiative electron transitions shown. The two spin transitions for 0 ↔ +1 and -½ ↔ +½ are indicated with 1 and ½ connections, respectively. **(b)** and **(c)** display experimental (circles) and modelled (dashed line) values of the reference, signal and contrast of a $T_1$ relaxation measurement on the 0 ↔ +1 and -½ ↔ +½ transitions respectively.*

To determine appropriate transition rates, experimental data was compared to corresponding simulated measurements. $T_1$ relaxation measurements were performed according to the pulse sequence shown in the Figure 2e inset, with π-pulses resonant with either the 0 ↔ +1 or -½ ↔ +½ transitions. The normalized photon counts within the reference and signal gates, along with the resulting ODMR contrast for each transition, are shown in Figure 3b and 3c, respectively. To emulate the experimental procedure, optical pumping was applied until the modelled system reached a steady state. The ground state population was subsequently evaluated as a function of dark time ($\tau$), providing a theoretical analogue of photon counts during readout. The excited state lifetime of this spin complex was measured to be 2.5 ns (Fig. 1d). We assumed $\kappa_r$ and the combined intersystem crossing rates ($ISC_{23} + ISC_{24}$) to be equal, yielding a quantum efficiency of 50%, consistent with reported values for hBN QEs[37].

The rates used to best reproduce the experimental data are listed in Table 1. The steady state fluorescence, observed at short dark times, reaches only 90% of the peak value measured at long dark intervals. Attempts to adjust the quantum efficiency did not yield accurate modelling unless $\kappa_p$ was set to a relatively low value. This suggests that, under experimental conditions, the optical excitation process is inefficient. It must be noted that the experimental and modelled signal and reference curves diverge at longer dark intervals. However, the modelled contrast remains consistent with the experimental data. This behaviour may indicate the presence of an additional, longer lived metastable state that does not carry any spin information. Nevertheless, the constructed spin complex model accurately describes the intricate photodynamics of cascading metastable states with separately addressable spin transitions, providing important insights into the hBN spin complex's behaviour. More information on the spin dependent rate model can be found in the Supplementary Information.

***Table 1.*** *Electronic transition rates used to model the $T_1$ relaxation dynamics. Rates are shown in $s^{-1}$.*

| Transition Rates ($s^{-1}$) | | | | | | | | | |
|---|---|---|---|---|---|---|---|---|---|
| $\kappa_p$ | $\kappa_r$ | $ISC_{23}$ | $ISC_{24}$ | $\kappa_{35}$ | $\kappa_{46}$ | $\kappa_{51}$ | $\kappa_{61}$ | $\omega_{34}$ | $\omega_{56}$ |
| $3.2 \times 10^4$ | $2 \times 10^8$ | $2 \times 10^6$ | $2 \times 10^8$ | $5 \times 10^7$ | $2 \times 10^5$ | $1 \times 10^8$ | $4 \times 10^4$ | $5.6 \times 10^4$ | $5.6 \times 10^4$ |

To gain further insight, we also investigate temperature dependent behaviour of the spin complex emitters from room to cryogenic temperature (300 K ~ 20 K). Figure 4 shows temperature dependent measurement of PL spectrum as well as ODMR measurement from a spin complex emitter in hBN. The PL characteristics of this emitter including autocorrelation measurement are summarized in Figure SI3. Temperature-dependent PL spectra in Figure 4a reveal a gradual blue shift of the ZPL with decreasing temperature, accompanied by a reduction in linewidth. The lattice contraction during the cool down modifies the ZPL emission energy while narrowing of the linewidth arises from reduced electron-phonon coupling at lower temperatures. The ZPL wavelengths and the full width at half maximum (FWHM) values are plotted in Figure 4b, showing a decrease in both parameters with lowering temperature. The total blue shift of approximately 2 nm over the 300 - 17 K range is recorded confirming the thermal effect on this quantum emitter.

Temperature-dependent ODMR spectra of an emitter with ZPL at 703 nm (see Figure SI4 for its optical characterisation) are shown in Figure 4c, with transitions -½ ↔ +½ and 0 ↔ +1 labeled above the corresponding peaks and resonances centred at 1.39 GHz and 2.09 GHz, respectively. Change in contrast and line shape of the -½ ↔ +½ transition is attributed to variations in the applied microwave power, which affects linewidths of ODMR transitions[10]. Across the investigated temperature range, the resonance frequencies for both transitions remained at similar frequencies. The -½ ↔ +½ spin transition is not expected to shift with temperature due to the weak dipolar coupling between the individual spins. The -½ ↔ +½ transitions are accompanied by the second harmonic from double quantum transition in spin complex emitters which is well resolved at lower temperatures in our measurement.

To account for any variation in the applied external magnetic field on the sample due to thermal shift, ODMR spectra was first adjusted to make the –½ ↔ +½ transitions overlap. Then, the position of the 0 ↔ +1 transition transition was recorded at varying temperatures. From Figure 4d, a maximum shift of 2.3 ± 0.9 MHz was observed for the 0 ↔ +1 transition, demonstrating a very weak interaction between the zero-field splitting (ZFS) of the triplet state and temperature-induced changes to the hBN. Additionally, in Figure SI6, a shift of 10.9 ± 1.2 MHz was observed for an emitter with ZPL at 721 nm. For comparison, the change in the ZFS for the $V_B^-$ center in hBN is ~195 MHz at 10 K[17]. This behaviour is promising for applying spin-complex emitters in hBN for robust, temperature-stable magnetometry.

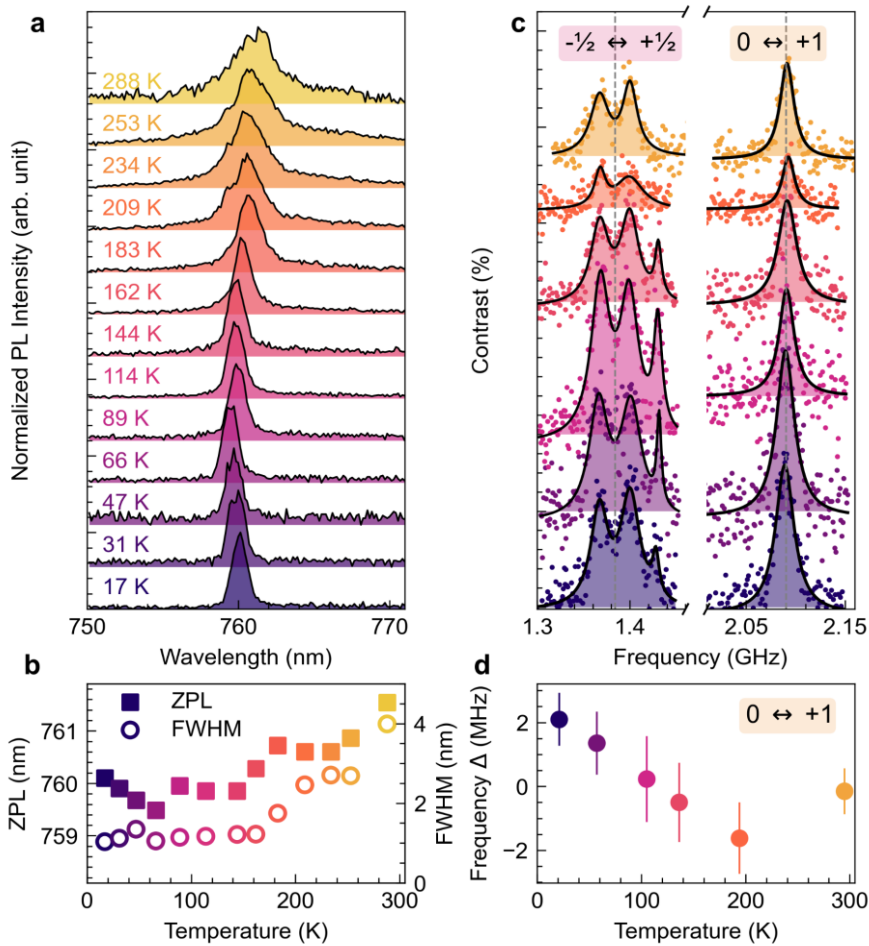

*Figure 4. Temperature dependent ODMR of the spin complex. (a) PL spectra of a spin-active emitter with ZPL at 760 nm. (b) Extracted ZPL wavelength and FWHM values of the acquired spectra at corresponding temperatures. (c) ODMR spectra at varying temperatures of a spin-active emitter with ZPL at 703 nm. (d) Change in resonance frequency position of the 0 ↔ +1 transition at corresponding temperatures.*

To further characterize the effect of temperature on the spin-dependent recombination of the spin complex, pulsed ODMR measurements were performed at room (295 K) and cryogenic temperatures (20 K) for both the 0 ↔ +1 and –½ ↔ +½ transitions. The photoluminescence, CW-

ODMR and Rabi oscillations of the emitter used for these measurements are shown in Figures SI5-8. The corresponding π- and π/2-pulse timings are extracted from the Rabi oscillations, allowing for $T_1$ relaxation and spin echo measurements to be performed at both 295 K and 20 K.

The results for the 0 ↔ +1 and -½ ↔ +½ transitions at both temperatures are shown in Figure 5a,b, respectively. For the 0 ↔ +1 transition (Figure 5a), $T_{rise}$ increases from 180 ns at 295 K to 633 ns at 20 K, while $T_1$ only increases slightly from 18 μs at 295 K to 20 μs at 20 K. For the -½ ↔ +½ transition however the changes are more pronounced, with $T_{rise}$ increasing by an order of magnitude from 259 ns at 295 K to 3104 ns at 20 K, and $T_1$ increasing by more than double from 13 μs to 30 μs.

The greater increase in $T_1$ for the -½ ↔ +½ transition compared to the 0 ↔ +1 transition further support the spin model of spin complex emitters. As the spin-pair is delocalised over two defects it is likely more sensitive to spin-phonon interactions, meaning it has a greater dependence on temperature than that of the triplet state localized on a single defect site. Furthermore, the increase in $T_{rise}$ for both transitions at cryogenic temperatures suggests that the lifetimes of the short lived metastable state is enhanced. Additionally, the order of magnitude increase in $T_{rise}$ for the -½ ↔ +½ transition indicates that the relaxation of the charge–transfer state is potentially thermally assisted.

Similarly, the results for the spin echo was measured for the 0 ↔ +1 and -½ ↔ +½ transitions at 295 K and 20 K shown in Figure SI8a,b, respectively. Both transitions display an increase in $T_2$ when comparing 295 K and 20 K, signifying an enhancement in the spin coherence time with decreasing temperature. As the in-plane distance between atoms in hBN expands with decreasing temperature[17,38], the dipolar coupling between surrounding spins is decreased, leading to a longer coherence time. Notably, the observed increase in $T_2$ is greater for the -½ ↔ +½ transition, increasing from 40 ns to 52 ns between 295 K and 20 K. This behaviour aligns with the proposed spin pair model[11], in which the -½ ↔ +½ transition is anticipated to be more sensitive to variations in the surrounding spin environment from changes in the lattice due to its weak dipolar coupling. Similar increases in coherence times have been observed for spin active defects in diamond and in silicon carbide (SiC)[39–44], where NV⁻ centers in diamond exhibited a 375 ns increase (from 300 to 20 K)[44] and divacancies in SiC showed a 20 μs increase (from 300 K to 25 K)[39].

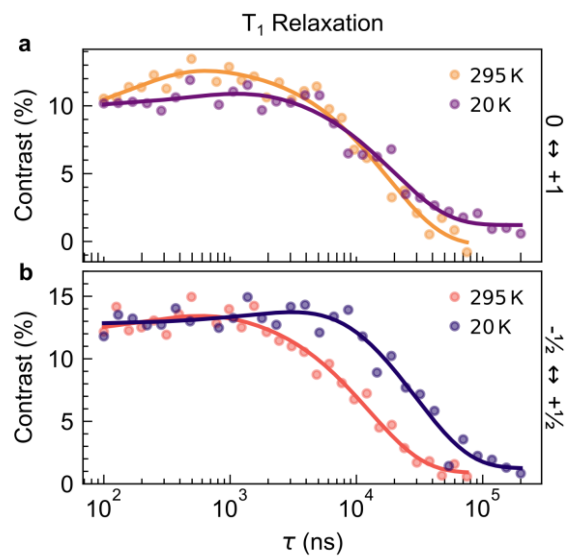

*Figure 5. Room and cryogenic temperature $T_1$ relaxation measurements. (a) and (b) show the $T_1$ measurements performed on the 0 ↔ +1 and -½ ↔ +½ transitions, respectively, at 295 K and 20 K. The*

*solid lines in **(a)** and **(b)** are bi-exponential fits fitted to the data points and the fitting parameters are detailed in Table 2.*

***Table 2.** Bi-exponential fitting parameters for the $T_1$ relaxation measurement shown in Figure 5.*

| Transition | Temperature (K) | $T_1$ Relaxation | |
|---|---|---|---|
| | | $T_{rise}$ (ns) | $T_1$ ($\mu s$) |
| $0 \leftrightarrow +1$ | 295 | 180 | 18 |
| | 20 | 633 | 20 |
| $-½ \leftrightarrow +½$ | 295 | 259 | 13 |
| | 20 | 3104 | 29 |

**CONCLUSION**

To conclude, we performed a detailed photophysical analysis and temperature dependence studies of spin complex defects in hBN. By analyzing photodynamics of the spin complex defects, we first confirm that the spin dependent transitions exist as a metastable excited state. Additionally, we also verify that the population of the $S = 1$ triplet state first cascades through to the weakly coupled $S = ½$ spin pair prior to relaxation to the ground state. From temperature dependence measurements we found that the frequency $S = 1$ transition in spin complex defects are only weakly affected by changes in temperature. In addition, we show that while the rise, relaxation, and coherence times all increase with decreasing temperature for both transitions, the $S = ½$ transition exhibits a stronger temperature dependence. This suggests that the weakly coupled spin pair is more strongly influenced by the surrounding environment, providing an additional clue to the structure of spin complex defects in hBN. Whilst further work is required before single spin defects in hBN can be used for practical quantum sensing, our results support the proposed spin-pair model and provide additional phenomenological insights into the nature of the defect.

**METHODS**

PL and ODMR characterisations were carried out using home-built confocal microscope systems operating in reflection measurement and equipped with high numerical aperture (>0.7 NA) objectives. PL was collected into multimode fibre either for spectrometer imaging (Andor SR303I-Newton CCD) using 300 l/mm grating or for single photon detection with avalanche photodiodes (Excelitas APD). For correlation measurements, collected light passed through a 50:50 beamsplitter fiber with coincidence counts analysed by a counter module (PicoQuant PicoHarp 300). ODMR measurements used a NdFeB magnet (N35) positioned perpendicular to the sample surface. Microwaves were generated using a signal generator (AnaPico APSIN 4010) and delivered onto a sample with suspended copper wire. An amplifier (Minicircuits ZHL-16W-43-S+) was used to increase microwave power. Pulse ODMR measurements employed an acousto-optical modulator in the optical excitation path and a radiofrequency switch (Minicircuits ZYSWA-2-50DR+) in the microwave path, both controlled by a pulse generator (Swabian PulseStreamer). During room temperature measurements, samples were kept at ambient conditions, while cryogenic measurements were conducted using a closed-loop helium cryostat (Attocube attoDRY 800) at varying temperatures and pressures of approximately $10^{-5}$ mbarr.

## ACKNOWLEDGEMENTS

The authors acknowledge financial support from the Australian Research Council (CE200100010, FT220100053, and DP240103127), the Air Force Office of Scientific Research (FA2386-25-1-4044).

## REFERENCES

1. Aharonovich, I., Englund, D. & Toth, M. Solid-state single-photon emitters. *Nat. Photonics* **10**, 631–641 (2016).
2. Awschalom, D. D., Hanson, R., Wrachtrup, J. & Zhou, B. B. Quantum technologies with optically interfaced solid-state spins. *Nat. Photonics* **12**, 516–527 (2018).
3. Wolfowicz, G. *et al.* Quantum guidelines for solid-state spin defects. *Nat. Rev. Mater.* **6**, 906–925 (2021).
4. Atatüre, M., Englund, D., Vamivakas, N., Lee, S.-Y. & Wrachtrup, J. Material platforms for spin-based photonic quantum technologies. *Nat. Rev. Mater.* **3**, 38–51 (2018).
5. Widmann, M. *et al.* Coherent control of single spins in silicon carbide at room temperature. *Nat. Mater.* **14**, 164–168 (2015).
6. He, Y.-M. *et al.* Single quantum emitters in monolayer semiconductors. *Nat. Nanotechnol.* **10**, 497–502 (2015).
7. Stern, H. L. *et al.* Room-temperature optically detected magnetic resonance of single defects in hexagonal boron nitride. *Nat. Commun.* **13**, 618 (2022).
8. Scholten, S. C. *et al.* Multi-species optically addressable spin defects in a van der Waals material. *Nat. Commun.* **15**, 6727 (2024).
9. Stern, H. L. *et al.* A quantum coherent spin in hexagonal boron nitride at ambient conditions. *Nat. Mater.* **23**, 1379–1385 (2024).
10. Gao, X. *et al.* Single nuclear spin detection and control in a van der Waals material. *Nature* **643**, 943–949 (2025).
11. Robertson, I. O. *et al.* A charge transfer mechanism for optically addressable solid-state spin pairs. *Nat. Phys.* https://doi.org/10.1038/s41567-025-03091-5 (2025).
12. Whitefield, B. *et al.* Generation of narrowband quantum emitters in hBN with optically addressable spins.
13. Sierra, J. F., Fabian, J., Kawakami, R. K., Roche, S. & Valenzuela, S. O. Van der Waals heterostructures for spintronics and opto-spintronics. *Nat. Nanotechnol.* **16**, 856–868 (2021).
14. Moon, S. *et al.* Hexagonal Boron Nitride for Next-Generation Photonics and Electronics. *Adv. Mater.* **35**, 2204161 (2023).
15. Nonahal, M. *et al.* Engineering Quantum Nanophotonic Components from Hexagonal Boron Nitride. *Laser Photonics Rev.* **17**, 2300019 (2023).

16. Gottscholl, A. *et al.* Initialization and read-out of intrinsic spin defects in a van der Waals crystal at room temperature. *Nat. Mater.* **19**, 540–545 (2020).
17. Gottscholl, A. *et al.* Spin defects in hBN as promising temperature, pressure and magnetic field quantum sensors. *Nat. Commun.* **12**, 4480 (2021).
18. Healey, A. J. *et al.* Quantum microscopy with van der Waals heterostructures. *Nat. Phys.* **19**, 87–91 (2023).
19. Huang, M. *et al.* Wide field imaging of van der Waals ferromagnet Fe3GeTe2 by spin defects in hexagonal boron nitride. *Nat. Commun.* **13**, 5369 (2022).
20. Robertson, I. O. *et al.* Detection of Paramagnetic Spins with an Ultrathin van der Waals Quantum Sensor. *ACS Nano* **17**, 13408–13417 (2023).
21. Durand, A. *et al.* Optically Active Spin Defects in Few-Layer Thick Hexagonal Boron Nitride. *Phys. Rev. Lett.* **131**, 116902 (2023).
22. Chejanovsky, N. *et al.* Single-spin resonance in a van der Waals embedded paramagnetic defect. *Nat. Mater.* **20**, 1079–1084 (2021).
23. Guo, N.-J. *et al.* Coherent control of an ultrabright single spin in hexagonal boron nitride at room temperature. *Nat. Commun.* **14**, 2893 (2023).
24. Gao, X. *et al.* Nanotube spin defects for omnidirectional magnetic field sensing. *Nat. Commun.* **15**, 7697 (2024).
25. M. Gilardoni, C. *et al.* A single spin in hexagonal boron nitride for vectorial quantum magnetometry. *Nat. Commun.* **16**, 4947 (2025).
26. Luo, J., Geng, Y., Rana, F. & Fuchs, G. D. Room temperature optically detected magnetic resonance of single spins in GaN. *Nat. Mater.* **23**, 512-518 (2024).
27. Vaidya, S. *et al.* Coherent Spins in van der Waals Semiconductor $GeS_2$ at Ambient Conditions. *Nano Lett.* **25**, 14356–14362 (2025).
28. Liu, W. *et al.* Experimental Observation of Spin Defects in the van der Waals Material $GeS_2$. *Nano Lett.* **25**, 16330–16339 (2025).
29. Patel, R. N. *et al.* Probing the Optical Dynamics of Quantum Emitters in Hexagonal Boron Nitride. *PRX Quantum* **3**, 030331 (2022).
30. Steiner, M., Neumann, P., Beck, J., Jelezko, F. & Wrachtrup, J. Universal enhancement of the optical readout fidelity of single electron spins at nitrogen-vacancy centers in diamond. *Phys. Rev. B* **81**, 035205 (2010).
31. Baber, S. *et al.* Excited State Spectroscopy of Boron Vacancy Defects in Hexagonal Boron Nitride Using Time-Resolved Optically Detected Magnetic Resonance. *Nano Lett.* **22**, 461–467 (2022).
32. Singh, H. *et al.* Room-temperature quantum sensing with photoexcited triplet electrons in organic crystals. *Phys. Rev. Res.* **7**, 013192 (2025).
33. Doherty, M. W. *et al.* The nitrogen-vacancy colour centre in diamond. *Phys. Rep.* **528**, 1–45 (2013).
34. Whitefield, B., Toth, M., Aharonovich, I., Tetienne, J. & Kianinia, M. Magnetic Field Sensitivity Optimization of Negatively Charged Boron Vacancy Defects in hBN. *Adv. Quantum Technol.* **8**, 2300118 (2025).
35. Mena, A. *et al.* Room-Temperature Optically Detected Coherent Control of Molecular Spins. *Phys. Rev. Lett.* **133**, 120801 (2024).
36. Lee, S.-Y. *et al.* Readout and control of a single nuclear spin with a metastable electron spin ancilla. *Nat. Nanotechnol.* **8**, 487–492 (2013).
37. Nikolay, N. *et al.* Direct measurement of quantum efficiency of single-photon emitters in hexagonal boron nitride. *Optica* **6**, 1084 (2019).

38. Paszkowicz, W., Pelka, J. B., Knapp, M., Szyszko, T. & Podsiadlo, S. Lattice parameters and anisotropic thermal expansion of hexagonal boron nitride in the 10-297.5 K temperature range. *Appl. Phys. Mater. Sci. Process.* **75**, 431–435 (2002).

39. Lin, W.-X. *et al.* Temperature dependence of divacancy spin coherence in implanted silicon carbide. *Phys. Rev. B* **104**, 125305 (2021).

40. Embley, J. S. *et al.* Electron spin coherence of silicon vacancies in proton-irradiated 4*H*-SiC. *Phys. Rev. B* **95**, 045206 (2017).

41. Falk, A. L. *et al.* Polytype control of spin qubits in silicon carbide. *Nat. Commun.* **4**, 1819 (2013).

42. Lin, S. *et al.* Temperature-dependent coherence properties of NV ensemble in diamond up to 600 K. *Phys. Rev. B* **104**, 155430 (2021).

43. Bar-Gill, N., Pham, L. M., Jarmola, A., Budker, D. & Walsworth, R. L. Solid-state electronic spin coherence time approaching one second. *Nat. Commun.* **4**, 1743 (2013).

44. Takahashi, S., Hanson, R., Van Tol, J., Sherwin, M. S. & Awschalom, D. D. Quenching Spin Decoherence in Diamond through Spin Bath Polarization. *Phys. Rev. Lett.* **101**, 047601 (2008).